\documentclass[journal]{IEEEtran}
\usepackage{graphicx}
\usepackage{placeins}
\usepackage{float}
\usepackage{tabularx,colortbl}
\usepackage{amsthm}
\usepackage{amsmath}
\usepackage{mathtools}
\usepackage{amssymb}
\usepackage{algorithm}
\usepackage{verbatim}
\usepackage{url}
\usepackage{cases}
\usepackage{amsmath}
\usepackage{amsfonts}
\usepackage{amssymb}
\usepackage{verbatim}
\usepackage{multirow}
\usepackage{graphicx}
\usepackage{epstopdf}
\usepackage{subfigure}
\usepackage{graphics}
\usepackage{color}

\IEEEoverridecommandlockouts

\begin{document}
\title{WINDOW: Wideband Demodulator for Optical Waveforms}
\author{Omri Lev, Tal Wiener, Deborah Cohen, \emph{Student IEEE}, Yonina C. Eldar, \emph{Fellow IEEE}
\thanks{
This project has received funding from the European Union's Horizon 2020 research and innovation program under grant agreement No. 646804-ERC-COG-BNYQ, and from the Israel Science Foundation under Grant no. 335/14. Deborah Cohen is grateful to the Azrieli Foundation for the award of an Azrieli Fellowship.}}

\maketitle

\begin{abstract}
Optical communication systems, which operate at very high rates, are often limited by the sampling rate bottleneck. The optical wideband regime may exceed analog to digital converters (ADCs) front-end bandwidth. Multi-channel sampling approaches, such as multicoset or interleaved ADCs, have been proposed to sample the wideband signal using several channels. Each channel samples below the Nyquist rate such that the overall sampling rate is preserved. However, this scheme suffers from two practical limitations that make its implementation difficult. First, the inherent anti-aliasing filter of the samplers distorts the wideband signal. Second, it requires accurate time shifts on the order of the signal's Nyquist rate, which are challenging to maintain. In this work, we propose an alternative multi-channel sampling scheme, the wideband demodulator for optical waveforms (WINDOW), based on analog RF demodulation, where each channel aliases the spectrum using a periodic mixing function before integration and sampling. We show that intentionally using the inherent ADC filter to perform integration increases the signal to noise ratio (SNR). We demonstrate both theoretically and through numerical experiments that our system outperforms multicoset in terms of signal recovery and symbol estimation in the presence of both thermal and quantization noise but is slightly less robust to timing jitter.
\end{abstract}

\section{Introduction}

Modern optical communication systems, used in many telecommunication companies such as telephone, Internet and cable television, operate at very high data rates, with typical values of $20$ GHz \cite{Laperle}. Nyquist rates of such signals may exceed the specifications of today's high-end analog-to-digital converters (ADCs) by orders of magnitude. High rate ADCs typically have low resolution, down to even 1 bit, due to hardware and cost limitations, which effectively reduces the data rate. This trade-off between bandwidth and resolution is the main limitation to increasing the rate in communication systems, and alternative sampling schemes have been proposed to overcome it \cite{Azadet}.

The most popular solution, referred to as multicoset sampling or interleaved ADCs \cite{Laperle, Bresler, Mishali_multicoset, SamplingBook}, adopts a multi-channel approach. Instead of implementing a single ADC at high rate $f_{\text{Nyq}}$, interleaved ADCs use $M$ devices sampling at the lower rate $f_{\text{Nyq}}/M$ with appropriate time shifts, thus benefiting from the higher resolution of low rate ADCs. Each channel processes a delayed version of the original signal, so that combining the low rate samples results in the Nyquist samples of the original high rate signal, with higher resolution than that obtained using a high rate ADC. The number of channels $M$ governs the trade-off between hardware complexity and sampling rate per channel. However, time interleaving has two fundamental shortcomings. First, practical ADCs introduce an inherent bandwidth limitation, modeled by an anti-aliasing low-pass filter (LPF), which distorts the samples \cite{ADC_bandwidth}. Multicoset thus requires low rate samplers with high analog front-end bandwidth. As a consequence, implementing multicoset for optical wideband signals requires the design of specialized fine-tuned ADC circuits that meet the high analog bandwidth requirements, which is challenging with today’s technology \cite{ADC_bandwidth, Mishali_theory}. Second, maintaining accurate time shifts on the order of the Nyquist rate is difficult to implement \cite{Razavi}.

In this paper, we present a wideband demodulator for optical waveforms (WINDOW), an alternative sampling system, inspired by the random demodulator (RD) from \cite{RandomDemodulator2} and the modulated wideband converter (MWC) \cite{Mishali_theory}. We combine the multi-channel approach of the previous technique, to reduce the sampling rate of each ADC, with the advantages of analog RF demodulation. To circumvent the analog bandwidth issue in the ADCs, an RF front-end mixes the input signal with periodic waveforms. This operation imitates the effect of delayed undersampling used in the multicoset scheme and results in folding the spectrum to baseband with different weights for each frequency interval. The mixed signal is then integrated over a fixed period, similar to the RD approach, and sampled at a low rate. We then show how the Nyquist rate samples can be digitally recovered from the resulting samples. This system allows the transfer of high rate signals with today's available hardware technology.

Next, we analyze the effect of two sources of noise on the performance of signal recovery of WINDOW: thermal and quantization noise. We theoretically compare the impact of noise on both our system and multicoset. We show that the integration operation in WINDOW increases signal to noise ratio (SNR) with respect to the thermal noise whereas the resulting quantization noise is higher. Since the latter is typically smaller than thermal noise by orders of magnitude, our system is more robust to noise overall. This observation is corroborated by simulations. We also consider the effect of timing jitter on both systems, whose impact is experimentally shown to be higher on our system than on multicoset. After equalization, we estimate the transmitted symbols from both sampling systems and our recovery error is shown to be lower than that of multicoset in the presence of both jitter and noise. This is due to the fact that, in typical settings, the increase in thermal SNR is greater than the impact of quantization noise and timing jitter.

This paper is organized as follows. In Section~\ref{sec:back}, we describe some fundamentals of optical communication systems. Section~\ref{sec:multi} reviews the multicoset system and its limitations. In Section~\ref{sec:algo}, we present the WINDOW system, composed of an analog sampling front-end followed by digital recovery. Section~\ref{sec:noise} investigates its robustness to thermal and quantization noise as well as timing jitter with respect to multicoset sampling. Numerical experiments are shown in Section~\ref{sec:exp}.

\section{Formulation And Background}
\label{sec:back}

\begin{figure}
\begin{center}
\includegraphics[width=0.9\columnwidth]{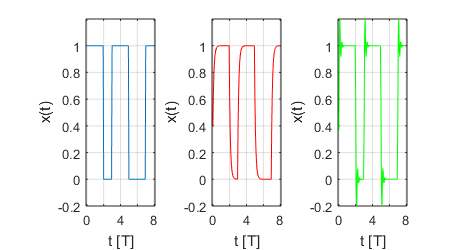}
\caption{On-off keying (2 PAM) with different pulse shapes: (a) ideal rectangular pulse, (b) LED and (c) laser.}
\label{fig:pamsig}
\end{center}
\end{figure}

Optical communication relies on transferring high-rate signals between two end points, typically using low order pulse amplitude modulation (PAM) signals. Such modulated signals are characterized by sets of discrete voltage levels, whose number usually sets the communication system resolution. PAM signals are expressed as
\begin{equation} \label{eq:pam}
x(t)=\sum_{k}a_{k} g_T(t-kT),
\end{equation}
where $1/T$ is the symbol rate, the set of amplitudes $\{a_k\}$ are referred to as symbols and $g_T(t)$ is the transmitted pulse shape \cite{CommBook}. The number of voltage levels is equal to $2^{B_{\text{in}}}$, where $B_{\text{in}}$ is the number of bits per symbol. According to the Nyquist criterion to avoid inter-symbol interference (ISI), it is required that $g_T(nT)=\delta[n]$. While ideally $g_T(t)$ is a square window function, in practice it has finite bandwidth, governed by the hardware technology. Today, most optical communication systems generate PAM signals using LED and laser technologies \cite{AgrawalBook}. For LED generators, the pulse shape is modeled as \cite{IvanBook}:
\begin{equation}
\label{eq:led}
g_T(t) = 1-e^{-\frac{t-kT}{\tau}},
\end{equation}
where $\tau$ denotes the LED time constant. Other systems use Gaussian pulses \cite{Tabrikian_optics}, such that
\begin{equation}
\label{eq:gauss}
g_T(t)=e^{-\frac{1}{2}(\frac{t}{T_0})^2},
\end{equation}
where $T_0$ is the pulse width.
The ideal rectangular PAM and corresponding LED and laser generated PAM for on-off keying or 2 PAM are shown in Fig.~\ref{fig:pamsig}. This corresponds to symbols $a_k$ with $B_{\text{in}} =1$ bit.

Optical communication systems use optical fiber links for signal transferring. Typical optical fibers include a transparent core surrounded by a transparent cladding material with a lower index of refraction. Light is kept inside the core from the total internal reflection which causes the fiber to act as a waveguide. As it travels through the transmission medium, the signal is affected by two main phenomena: attenuation and refraction \cite{IvanOFDM}.
First, the intensity of the light beam, or signal, is exponentially reduced with respect to the link length $L_0$. The received intensity $I_{\text{out}}$ is related to the input intensity $I_{\text{in}}$ as
\begin{equation}
I_{\text{out}}=I_{\text{in}}e^{-\alpha L_0}
\end{equation}
where $\alpha$ is the attenuation factor with typical value of $0.2\frac{\text{dB}}{\text{km}}$. 

In addition, the refractive index of fibers varies slightly with the frequency of light, and light sources are not perfectly monochromatic. Modulation of the light source also slightly widens the frequency band of the transmitted light. This results in the fact that different frequencies of light, transmitted over long distances and at high modulation speeds, arrive at different times at the receiver. Mathematically, the received signal in the frequency domain is given by the relation
\begin{equation} \label{eq:ref}
X_{out}(f)=X_{in}(f) e^{j\beta_{0}} e^{jL\beta_{1}f} e^{jL_0\frac{\beta_2}{2}f^2},
\end{equation}
where $\beta_0$, $\beta_1$ and $\beta_2$ are parameters that depend on the fiber material, signal frequency and bandwidth that determine the refractive index of the light.
While the constant and linear elements are connected to constant phase shift and time shift, the quadratic element is equivalent to convolving the time domain signal with a complex Gaussian. This causes pulse widening and ISI between consecutive symbols that severely affect the received signal. The received signal thus follows the PAM model (\ref{eq:pam}), while exhibiting a different pulse shape $g(t)$ than that of the transmitted signal.

The two above phenomena strongly depend on the fiber length, which leads to a distinction between short and long range communication systems \cite{Azadet}. In short range communication, both distortions can be neglected and the received pulse is essentially equal to the transmitted one. In this case, there is no ISI and since no equalization is required, the signal is decoded using an optical slicer.
In long range communication, the signal is severely affected by the channel, and the original transmitted data cannot be decoded from a simple slicer. In this case, we need to equalize the received signal, which requires sampling at the symbol rate $1/T$. We refer to this rate as the Nyquist rate for clarity below, although it is not necessarily the Nyquist rate of $g(t)$. In practice, the signal is typically oversampled by 2 to 8 times the Nyquist rate. In this paper, we therefore focus on the issue posed by long range applications in terms of sampling rate and present a multi-channel system that allows to use low rate ADCs.

\section{Multicoset Sampling with Practical ADCs}
\label{sec:multi}

\begin{figure}
\includegraphics[width=1\columnwidth]{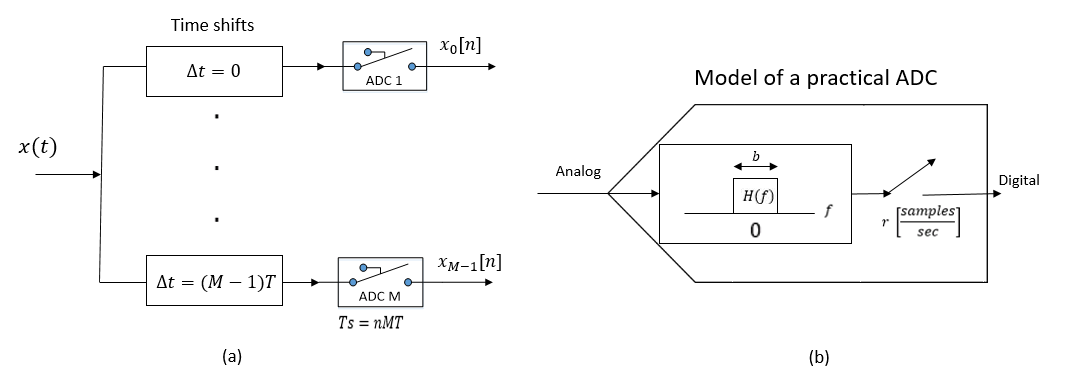}
\caption{(a) Schematic implementation of multicoset sampling, (b) Practical ADC front-end modeled as a LPF with bandwidth $b$ preceding the uniform sampling at rate $r$ samples/s \cite{Mishali_multicoset}.}
\label{fig:multico}
\end{figure}

A popular multi-channel sampling scheme for high rate signals is multicoset sampling, or interleaved ADCs. This approach samples $x(t)$ on a periodic nonuniform grid which is a subset of the Nyquist grid. Formally, denote by $x(nT)$ the sequence of samples taken at the Nyquist rate $T$. Multicoset samples consist of $M$ uniform sequences, called cosets, sampled at rate $1/(MT)$ with the $i$th coset defined by 
\begin{equation}
x_{i}[n]=x(nMT+iT), \qquad n\in \mathbb{Z},
\end{equation}
for $0 \leq i < M$.
Combining all $M$ cosets is equivalent to Nyquist rate sampling. Note that multicoset sampling has also been proposed in the context of sub-Nyquist sampling of multiband signals \cite{Mishali_multicoset}. There, only a fraction of the cosets are considered and the frequency sparse signal is recovered via sparse recovery \cite{SamplingBook} techniques. A possible implementation of multicoset is depicted in Fig.~\ref{fig:multico}.

Although this scheme seems relatively simple and straightforward, if suffers from two main drawbacks. First, practical ADCs introduce an inherent bandwidth limitation, crucial for high rate inputs, which distorts the samples. A uniform ADC attempts to output pointwise samples of the input. The design process and manufacturing technology result in an additional property, termed analog (full-power) bandwidth, which determines the maximal frequency handled by the device. The distortion mechanism can be modeled as a preceding LPF, as shown in Fig.~\ref{fig:multico}. Any spectral content beyond this frequency is attenuated and distorted. The ratio between the maximal frequency and sampling rate affects the complexity of the ADC circuit design \cite{ADC_bandwidth, Mishali_theory}, rendering the multicoset strategy challenging to implement.

The second issue arises from the time shift elements. Maintaining accurate time delays between the ADCs on the order of the Nyquist interval is difficult. Any uncertainty in these delays influences the recovery from the sampled sequences. A variety of different algorithms have been proposed in the literature in order to compensate for timing mismatches \cite{Razavi}. However, this adds substantial complexity to the receiver. These two difficulties limit multicoset performance, requiring the design of specific ADC circuits and complex digital recovery algorithms. Our solution adopts the multi-channel approach of multicoset sampling and combines it with RF analog demodulation, avoiding the practical limitations described above.

\begin{figure*}[h!]
	\begin{center}
		\includegraphics[width=1\textwidth]{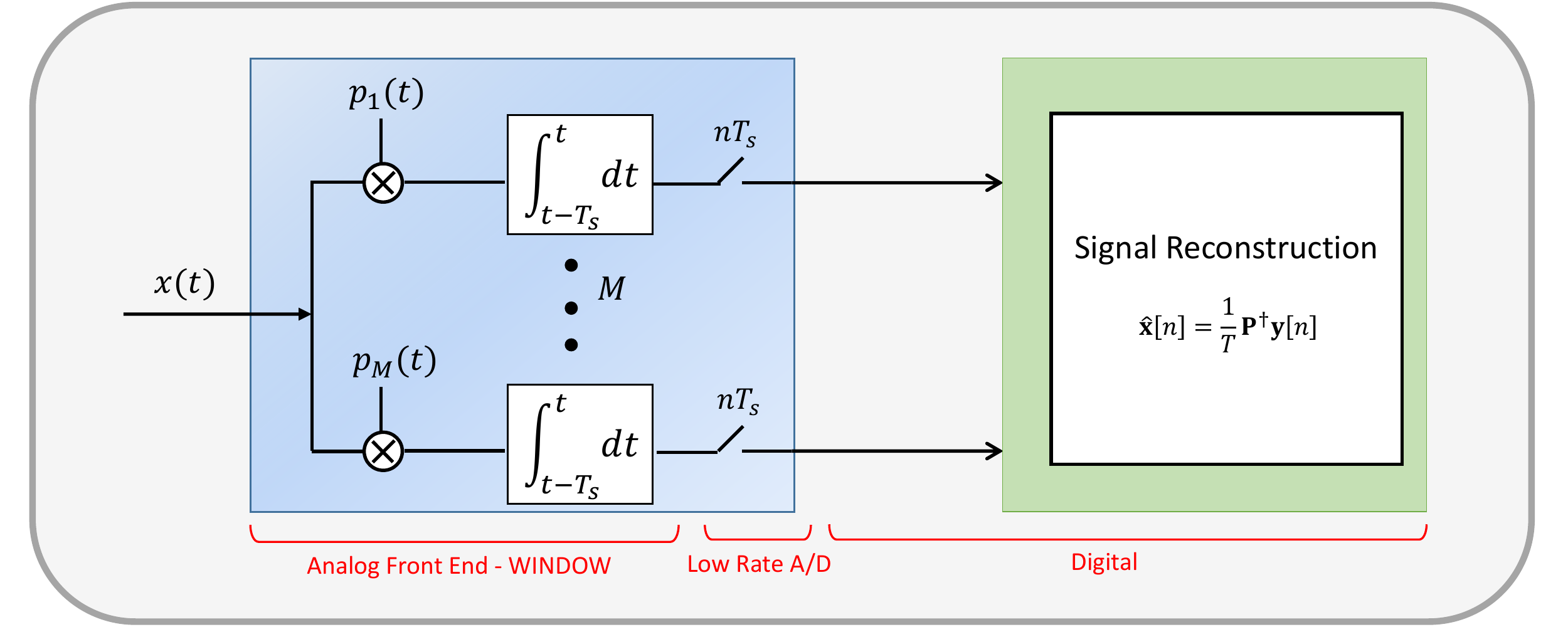}
		\caption{Schematic implementation of the WINDOW analog sampling front-end and digital signal recovery from low rate samples. In each channel, the input signal $x(t)$ is mixed with a periodic function $p_m(t)$ with period $T_s$, integrated over the period $T_s$ and sampled at the low rate $1/T_s$. The Nyquist samples $\mathbf{x}[n]$ are then recovered using (\ref{eq:solvex}).}
		\label{fig:HighLevel}
	\end{center}
\end{figure*}

\section{WINDOW System}
\label{sec:algo}
\subsection{Noiseless Signal Recovery}

Our solution is inspired by the RD \cite{RandomDemodulator2} and MWC \cite{Mishali_theory} proposed in the context of sparse sampling and recovery. 
Here, since we do not have any sparsity assumption on the signal, we use a multi-channel scheme in the Nyquist regime. We consider a received PAM signal with pulse shape $g_T(t)$ that can be rectangular, Gaussian or generated from LED or laser technologies, transmitted through a refractive channel (\ref{eq:ref}), creating ISI. Denote by $g(t)$ the pulse shape of the received signal $x(t)$. The signal is split and fed to $M$ channels. In each channel, $x(t)$ is first multiplied
by a unique periodic mixing function $p_m(t)$ with period $T_s$, whose basic period is given by 
\begin{equation}
p_{m}(t)=\sum_{l=0}^{L-1} p_{ml} u(t-lT),
\end{equation}
where $L=T_s/T$ is the mixing sequence's length, $\left\{ p_{ml}\right\}_{l=0}^{L-1} $ are random sequences alternating between the $\pm1$ levels and $u(t)$ is the rectangular function supported on $[0,T]$.

The resulting mixed signal,
\begin{equation}
\tilde{x}_m(t)=x(t) p_m(t),
\end{equation}
is then sampled at the low rate $1/T_s$ using an integrate and dump operation. The $n$th sample from the $m$th channel is given by 
\begin{eqnarray}
y_{m}\left[n\right] &=&\int_{(n-1)T_s}^{nT_s}\tilde{x}_m (t)\mathrm{d}t \nonumber \\
&=& \int_{(n-1)T_s}^{nT_s} x(t) \sum_{l=0}^{L-1} p_{ml} u(t-lT-(n-1)T_s) \mathrm{d}t \nonumber \\
&=& \sum_{l=0}^{L-1} p_{ml} \int_{(n-1)T_s+lT}^{(n-1)T_s+(l+1)T} x(t) \mathrm{d}t \nonumber \\
&\simeq&  T \sum_{l=1}^L x_{l}[n] p_{ml}.
\label{eq:samples}
\end{eqnarray}
Here, $x_{l}[n]=x((n-1)T_s+lT)$ are the original Nyquist rate samples of $x(t)$ in the $n$th period. The approximation in the last equation comes from the fact that, point-wise Nyquist samples are almost equal to the signal average over the corresponding Nyquist interval, obtained by the integrate and dump operation.
In practice, the demodulation and integration approach overcomes the bandwidth and timing limitations of multicoset sampling.

For our purposes, it will be convenient to write (\ref{eq:samples}) in matrix form as
\begin{equation} \label{eq:syst}
\mathbf{y}[n] = T \mathbf{P} \mathbf{x}[n], \quad n \in \mathbb{Z},
\end{equation}
where $\mathbf{y}[n]$ is the known $M \times 1$ vector that concatenates the samples $y_m[n]$ from each channel. The unknown vector $\mathbf{x}[n]$ contains the $L$ Nyquist rate samples $x_l[n]$ and the $M \times L$ matrix $\bf P$ is composed of the mixing sequences, such that the $(m,l)$th entry of $\bf P$ is given by $\mathbf{P}_{ml}=p_{ml}$.
Therefore, if $M \geq L$ and $\bf P$ is left-invertible, then the Nyquist samples $x_l[n]$ can be recovered from (\ref{eq:syst}), as
\begin{equation} \label{eq:solvex}
\mathbf{\hat{x}}[n]=\frac{1}{T} \mathbf{P}^{\dagger} \mathbf{y}[n], \quad n \in \mathbb{Z}.
\end{equation}
We can therefore reconstruct the Nyquist samples $x_{l}[n]$ by using $M$ samplers with sampling period $T_{s}=T/M$ by appropriate choice of mixing sequences. The analog WINDOW front-end and subsequent digital recovery are illustrated in Fig.~\ref{fig:HighLevel}.

\subsection{Channel Equalization}

We next recover the symbols from the reconstructed Nyquist samples of $x(t)$. Denote by $n_0$ the order of ISI and let $\kappa=\lfloor n_0/2 \rfloor$. For both schemes, we can relate the recovered Nyquist samples $\mathbf{\hat{z}}[n] = [\hat{x}[n-\kappa] \, \dots \, \hat{x}[n+\kappa]]^T$ selected from the corresponding entries of $\mathbf{\hat{x}}[n]$ to the symbol vector $\mathbf{a}[n]=[a[n-n_0+1] \, \dots \, a[n+n_0-1]]^T$ as
\begin{equation}
\mathbf{\hat{z}}[n]=\mathbf{H} \mathbf{a}[n] + \mathbf{n}[n],
\end{equation}
where the $(i,j)$th entry of the $n_0 \times 2n_0-1$ matrix $\bf H$ is given by
\begin{equation}
\mathbf{H}_{ij}= \left\{ 
\begin{array}{ll} g((j-i)T-\kappa T), & 0 \leq j-i \leq 2\kappa \\ 
0, & \text{else}, \end{array} \right.
\end{equation}
and $\mathbf{n}[n]$ is a noise vector. In the next section, we consider thermal and quantization noise and their impact on the WINDOW and multicoset systems. We adopt the popular linear equalization strategy in a minimum mean square error (MMSE) sense \cite{yeh_thesis, choi_book}, so that the $n$th symbol is estimated as
\begin{equation}
\hat{a}_n=\mathbf{c}^T_{\text{MMSE}} \mathbf{\hat{z}}[n],
\end{equation}
where
\begin{equation}
\label{eq:equa}
\mathbf{c}_{\text{MMSE}}=(\mathbf{HH}^T + \sigma_n^2\mathbf{I})^{-1}\mathbf{h}.
\end{equation}
Here, $\mathbf{h}$ is the vector of size $n_0$ defined by $\mathbf{h}= [g(-\kappa T) \, \dots \, g(\kappa T)]^T$ and $\sigma_n^2=\frac{\sigma_x^2}{\text{SNR}}$ is the noise variance, where $\sigma_x^2=\mathbb{E}x_m^2[n]$ and SNR are defined in (\ref{eq:snr}) for WINDOW and in (\ref{eq:snr_m}) for multicoset.


\section{Noise and Jitter Impact}
\label{sec:noise}

In the previous section, we considered recovery of $x(t)$ and its symbols $a_k$ under noiseless settings. We now analyze the impact of two types of noise on the performance of the WINDOW system, which we will next compare to the multicoset sampling scheme. We first consider the effect of additive measurement noise on the SNR before equalization and then turn to quantization noise.

\subsection{Thermal and Quantization Noise}

Assume that the received signal is corrupted by additive white noise $v(t)$ with spectral density $N_0$ independent from $x(t)$, so that the received signal can be written as
\begin{equation}
z(t)=x(t)+v(t).
\end{equation}
The samples from our system are then given by 
\begin{eqnarray}
y_{m}[n] &=& \int_{(n-1)T_s}^{nT_s}(x(\tau)+v(\tau)) p_{m}(\tau)d\tau \nonumber \\
&=& T \sum_{l=1}^L x_l[n]p_{ml}+v_{m}[n],
\end{eqnarray}
where $v_m[n]=\int_{(n-1)T_s}^{nT_s}v(\tau) p_{m}(\tau)d\tau$ is a random variable with variance
\begin{equation}
\label{eq:sigmav}
\sigma^2_v= \int_{(n-1)T_s}^{nT_s}N_0 p_{m}^{2}(\tau)d\tau=N_0T_s.
\end{equation}
We will show later on that the coherent integration applied here improves the SNR by a factor of $T$.

So far, we assumed the ADCs have infinite resolution. However, practical samplers have limited resolution governed by the number of bits $B_{s}$ and perform quantization on the samples. Denote by $Q(\cdot)$ the quantization function and by $q$ the quantization step. We consider a simple uniform quantizer where $q$ is chosen as the quotient between the dynamic range $\Delta$ of the signal at the input of the sampler, that is after integration, and the number of quantization levels $B_s$, namely
\begin{equation}
\label{eq:q}
q = \frac{\Delta}{2^{B_s}}.
\end{equation}
The resulting quantized samples are given by
\begin{eqnarray}
y_{m}[n]&=&Q \left( \int_{(n-1)T_s}^{nT_s}(x(\tau)+v(\tau)) p_{m}(\tau)d\tau\right) \nonumber \\
&\simeq& T \sum_{l=1}^L x_l[n] p_{ml}+v_m[n] +w_m[n], \label{eq:quant_samples}
\end{eqnarray}
where we used a white noise model for the quantization and the source signal.

The quantization noise $w_m[n]$ is then uniformly distributed between $\pm\frac{q}{2}$ and has variance $\sigma_w^2=q^2/12$. The symbols $a_k$ have $B_{\text{in}}$ bits of resolution and thus take values in the interval $[-2^{B_{\text{in}}-1},2^{B_{\text{in}}-1}]$. For a normalized pulse such that $\int_{0}^{T} g(t) \mathrm{d}t = T$, after modulation with $\pm 1$ sequences and integration over a period $T_s$, the signal after integration lies between $[-2^{B_{\text{in}}-1},2^{B_{\text{in}}-1}] T_{s}$. The dynamic range is then bounded by
\begin{equation}
\label{eq:delta}
\Delta \leq 2^{B_{\text{in}}} T_s=2^{B_{\text{in}}} LT.
\end{equation}
This equation suggests a trade-off between the sampling rate of each channel and the dynamic range required from the ADCs. Lower sampling rates, which lead to more channels, increase the integration time and in turn the dynamic range $\Delta$, thus requiring samplers with higher resolution for a fixed quantization step $q$.

Writing (\ref{eq:quant_samples}) in matrix form leads to 
\begin{equation} \label{eq:syst_noise}
\mathbf{y}[n] = T \mathbf{P} \mathbf{x}[n] + \mathbf{v}[n]+\mathbf{w}[n], \qquad n \in \mathbb{Z},
\end{equation}
where $\mathbf{v}[n]$ and $\mathbf{w}[n]$ are the thermal and quantization noise vectors that contain the $M$ samples $v_m[n]$ and $w_m[n]$, respectively.
Again, if $\bf P$ is left-invertible, then the estimator $\mathbf{\hat{x}}[n]$ defined in (\ref{eq:solvex}) from the noisy measurements (\ref{eq:syst_noise}) is related to the original signal samples $\mathbf{x}[n]$ by
\begin{equation}
\mathbf{\hat{x}}[n] = \mathbf{x}[n]+\mathbf{r}[n],
\end{equation}
where 
\begin{equation} \label{eq:r_def}
\mathbf{r}[n] \triangleq \frac{1}{T} \mathbf{P}^{\dagger}  (\mathbf{v}[n]+\mathbf{w}[n]).
\end{equation}

To avoid noise enhancement, we select orthogonal sets $\{p_m\}$, namely $\sum_{l=1}^{L}p_{m_1l} p_{m_2l}=0$, for $m_1 \neq m_2$, where $m_1$ and $m_2$ are channel indices. In particular, we propose to choose $\bf P$ as an $L \times L$ Hadamard matrix whose rows are orthogonal, leading to $\mathbf{P}^{-1}=\frac{1}{L} \mathbf{P}$, with entries equal to $\pm 1$ \cite{hadamard}.
The corresponding mixing functions $p_m(t)$ are the well known Walsh functions.
For the sake of SNR analysis, we assume that $x(t)$ is wide-sense stationary with variance $\sigma^2_x$. Since $x(t)$, $v(t)$ and the quantization noise are independent, $v_m[n]$ and $w_m[n]$ are independent as well. The resulting SNR of the reconstructed samples is given by
\begin{equation}
\label{eq:snr}
\text{SNR}=\frac{\mathbb{E}(x_m^2[n])}{\mathbb{E}(r_m^2[n])} = \frac{\sigma_x^2}{\frac{L}{T^2 L^2}(\sigma_v^2+\sigma_w^2)} = \frac{L T^2 \sigma_x^2}{N_0T_s + q^2/12}.
\end{equation}
Substituting (\ref{eq:q}) and (\ref{eq:delta}) into (\ref{eq:snr}), we have
\begin{equation} \label{eq:snrr}
\text{SNR}=\frac{\sigma_x^2}{N_0/T + 2^{2(B_{\text{in}}-B_s)}L/12}.
\end{equation}
We next compare (\ref{eq:snrr}) with the SNR obtained using multicoset sampling.

\subsection{Comparison with Multicoset Sampling}

In the presence of thermal and quantization noise, the samples obtained at the $m$th channel of the multicoset system are given by
\begin{equation}
y_m[n]=x_m[n]+v_m^c[n]+w_m^c[n].
\end{equation}
Since the multicoset system samples the input signal $x(t)$ without any integration, $v_m^c$ is a random variable with variance $N_0$ and $w_m^c$ is uniformly distributed between $\pm \frac{q}{2}$ where $q$ is defined in (\ref{eq:q}) with dynamic range $\Delta=2^{B_{\text{in}}}$.
The resulting SNR of the reconstructed samples is then
\begin{equation}
\label{eq:snr_m}
\text{SNR}=\frac{\mathbb{E}(x_m^2[n])}{\mathbb{E}((v_m^c)^2[n])+\mathbb{E}((w_m^c)^2[n])} = \frac{\sigma_x^2}{N_0+q^2/12}.
\end{equation}
Using (\ref{eq:q}) with $\Delta=2^{B_{\text{in}}}$, (\ref{eq:snr_m}) can be written as
\begin{equation}
\text{SNR}=\frac{\sigma_x^2}{N_{0}+2^{2(B_{\text{in}}-B_s)}/12}.
\end{equation}

The SNR expression (\ref{eq:snrr}) shows that, with respect to multicoset, the use of integration in the WINDOW system reduces the thermal noise power $N_0$ by a factor of $T$ while increasing the quantization noise $L$ times. Since quantization noise is typically lower than thermal noise by orders of magnitude, our system is more robust to noise overall. The anti-aliasing filter, which is one disadvantage of the multicoset system, is used to our advantage by performing intentional integration for noise averaging to increase SNR. In addition, WINDOW does not suffer from the ADC high analog bandwidth limitation since the signal is first integrated before sampling.

\subsection{Jitter Model}
\label{sec:jitter}

We next investigate the impact of timing jitter on both systems. Jitter is a random process that affects the performance of synchronized systems and is architecture dependent. In order to investigate the effect of jitter on WINDOW, we consider the simple hardware implementation, based on the RD \cite{RandomDemodulator2} and the MWC \cite{Mishali_theory}, shown in Fig~\ref{fig:HighLevel}. We observe that jitter can arise from the clock controlling the mixing sequences and the ADCs' clock. Since the latter is slower by orders of magnitude, precisely by a factor of $T_s/T$, it can be neglected. In the simulations, we will thus consider a unique source of jitter in the mixing process with unit interval (UI) equal to $T$. Note that the same clock controls all mixing sequences. Therefore, all channels suffer from identical timing jitter.

For multicoset sampling, we refer to the implementation presented in Fig.~\ref{fig:multico}. Here, jitter can arise from non-synchronization between the delays as well as from the ADCs' clock. Again, the latter is slower by a factor of $T_s/T$ and is neglected. Therefore, we will only consider a unique source of jitter in the time delays elements with UI equal to $T$ as in the WINDOW system.

Jitter is modeled as a random variable added to the clock instances. Denote by $t_{k+1}$ the $(k+1)$th clock rise. Following \cite{lee2001modeling}, we have
\begin{equation} \label{eq:jitter_m}
t_{k+1}=(k+1) T+ \sum_{n=0}^k \sqrt{C} w_n,
\end{equation}
where the process $\{w_n\}$ is an additive white Gaussian noise (AWGN) with zero expectation and unit variance. The constant $C$ is set in order to model a specified value of the cycle-to-cycle root mean square (RMS), given by
\begin{equation}
\text{RMS}=\sqrt{\text{var} \left(\sum_{n=0}^{T_s/T-1} \sqrt{C} w_n\right)} = \sqrt{C \frac{T_s}{T}}.
\end{equation}
The constant $C$ is chosen to express the RMS as a specified ratio, denoted by $p$, of the UI, that is $\text{RMS}=100p\cdot \text{UI}$. It follows that
\begin{equation}
C=\frac{T}{T_s}(p \text{UI})^2 = \frac{T^3}{T_s}p^2.
\end{equation}

In the simulations, the jitter is modeled as described by (\ref{eq:jitter_m}) in the mixing sequences' clock of WINDOW and that of the time delay elements in multicoset.
In WINDOW, time shifts translate to random cyclic shifts of each row of $\bf P$, affecting orthogonality between channels, which results in interference between different symbols. Thus, we expect that our system will be more affected by jitter than multicoset. 
The effect of both noise and jitter are illustrated in simulations in Section \ref{sec:exp}.


\section{Simulations}
\label{sec:exp}

\begin{figure} 
\begin{center} 
\includegraphics[width=1\columnwidth]{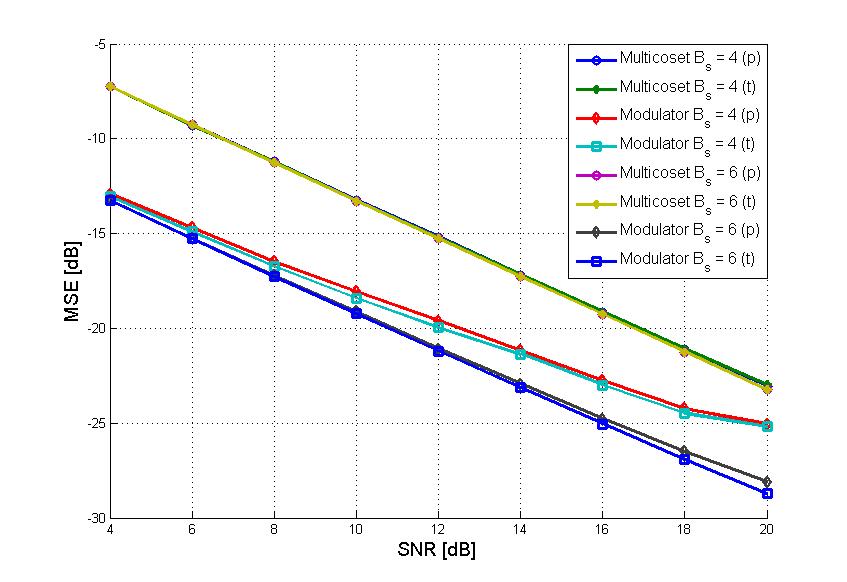}
\caption{Signal recovery error using on-off keying PAM for WINDOW and multicoset. In the legend, (e) denotes experimental and (t) refers to the theoretical bounds.}
\label{fig:no1}
\end{center}
\end{figure}

We now present some numerical experiments illustrating signal recovery and symbol estimation from low rate samples obtained using our proposed system in comparison with the multicoset scheme. We first generate a signal $x(t)$ produced using the LED technology with $g_T(t)$ defined in (\ref{eq:led}) and Nyquist rate $f_{\text{Nyq}}=1/T=20$GHz. We consider on-off keying modulation, namely $B_{\text{in}}=1$. Both WINDOW and multicoset are composed of $M=8$ channels, each sampling at $f_s=2.5$GHz. We simulate two sources of noise: thermal additive white Gaussian noise and quantization noise produced by finite resolution samplers setting a uniform quantization over the dynamic range of the sampled signal. Figure~\ref{fig:no1} presents the mean square error (MSE) between the true and recovered Nyquist samples with respect to SNR.
The results are shown for different values of number of bits $B_s$ of the ADCs, along with the theoretical bounds derived from (\ref{eq:snr}) and (\ref{eq:snr_m}). It can be seen that for multicoset, the experimental results coincide exactly with the theoretical bounds. For WINDOW, since the theoretical bound is computed for an ideal rectangular pulse, the experimental MSE is slightly higher than the bound. In addition, increasing $B_s$ has very little effect on the multicoset performance while it decreases the MSE of WINDOW, as expected.

\begin{figure} 
\begin{center} 
\includegraphics[width=1\columnwidth]{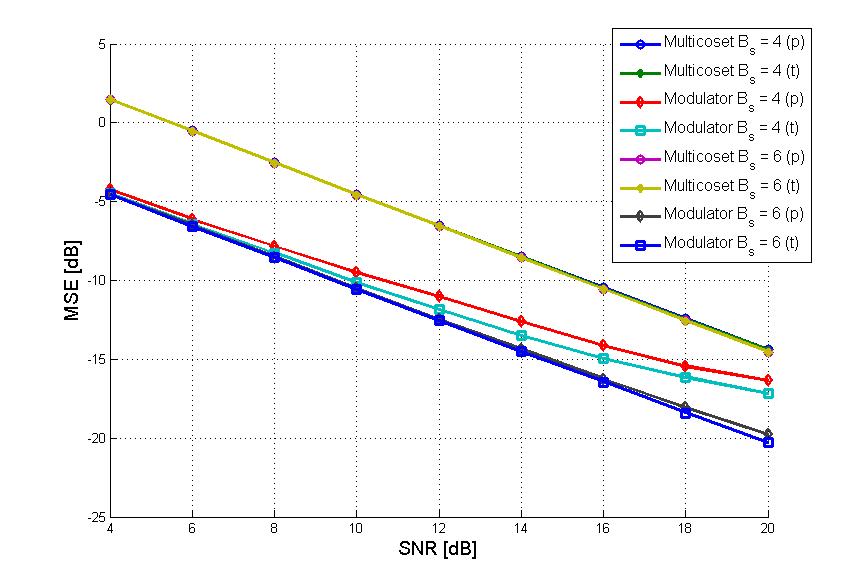}
\caption{Signal recovery error using 4 PAM and fiber length $L_0=140[\text{km}]$ for WINDOW and multicoset. In the legend, (e) denotes experimental and (t) refers to the theoretical bounds.}
\label{fig:new_mse}
\end{center}
\end{figure}

\begin{figure} 
\begin{center} 
\includegraphics[width=1\columnwidth]{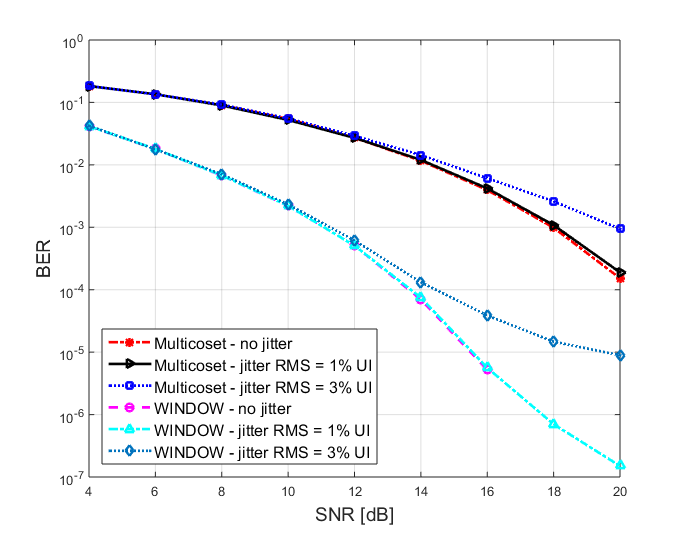}
\caption{Recovery error using on-off keying and fiber length $L_0=80[\text{km}]$ for WINDOW and multicoset.}
\label{fig:new_ber}
\end{center}
\end{figure}

In the previous simulation, the channel was almost not distorted. We now consider heavier distortion which results in ISI. To quantify the pulse widening, we consider a similar model to \cite{Tabrikian_optics}, where a Gaussian pulse (\ref{eq:gauss}) with $f_{\text{Nyq}}=10$GHz is transmitted. The pulse width $T_0$ is defined with respect to $T_{\text{fwhm}}=T/2$ which denotes the full width at half magnitude, so that $T_0=\frac{T_{\text{fwhm}}}{2\sqrt{2\ln(2)}}$. The channel response is given by $H(f)=e^{\frac{j\pi \lambda^2}{c}DL_0f^2}$, where $D=17\frac{\text{ps}}{\text{nm} \cdot \text{km}}$ is the chromatic dispersion parameter, $\lambda=1550\text{nm}$ is the wavelength, $L_0$ is the fiber length and $c$ is the speed of light. Here, $\beta_0=\beta_1=0$ and $\beta_2=2\pi \lambda^2D/c$.
The received pulse is then $g(t)=\frac{T_0}{\delta}e^{-\frac{1}{2}(\frac{t}{\delta})^2}$, where $\delta = \sqrt{T_0^2-j\frac{\lambda^2}{c}D\frac{L_0}{2\pi}}$ \cite{Tabrikian_optics}. Figure~\ref{fig:new_mse} shows the MSE between the true and recovered Nyquist samples for a Gaussian pulse with 4 PAM modulation after dispersion. We then perform equalization for both schemes, using (\ref{eq:equa}) for an ISI of $n_0=3$ symbols, as computed in \cite{Tabrikian_optics}. The resulting bit error rate (BER) for both WINDOW and multicoset is shown in the curves of Fig.~\ref{fig:new_ber} corresponding to no jitter.
There, the impact of timing jitter for both systems, as described in Section \ref{sec:jitter}, is demonstrated as well.

Our system outperforms multicoset in terms of signal recovery and symbol estimation, in the presence of thermal and quantization noises as well as timing jitter. For a given recovery error, our system can cope with lower SNR but is less robust to timing jitter errors.

\section{Conclusion}
In this work, we presented WINDOW, an alternative sampling scheme to multicoset for wideband communication signals, and in particular optical waveforms. Our system, a multi-channel RF demodulation system, overcomes multicoset's main practical drawbacks, resulting from low rate ADCs analog bandwidth. In addition, we shown both theoretically and through numerical experiments that our system reduces the SNR with respect to thermal noise by a factor equal to the symbol rate. In contrast, it increases the quantization noise and is less robust to jitter. Overall, under typical noise and jitter parameters, WINDOW outperforms multicoset since the gain in thermal SNR overcomes the smaller losses due to quantization noise and jitter.

\bibliographystyle{IEEEtran}
\bibliography{IEEEabrv,CR_ref}

\end{document}